\documentclass{aastex}
\usepackage{emulateapj5,apjfonts,amssymb}

\def\apj{ApJ}

\def\mnras{MNRAS}
\def\jcp{J. Comp. Phys.}
\def\lsim{~\raise0.3ex\hbox{$<$}\kern-0.75em{\lower0.65ex\hbox{$\sim$}}~}
\def\gsim{~\raise0.3ex\hbox{$>$}\kern-0.75em{\lower0.65ex\hbox{$\sim$}}~}

\slugcomment{published: ApJ 601, L55-L58, 2004}
\shorttitle{INTERSTELLAR TURBULENCE}
\shortauthors{KRITSUK \& NORMAN}

\begin{document}
\title{Scaling relations for turbulence in the multiphase interstellar medium}

\author{Alexei G. Kritsuk\footnotemark[1] and Michael L. Norman}
\affil{Department of Physics and Center for Astrophysics and Space Sciences, 
University of California at San Diego,\\
9500 Gilman Drive, La Jolla, CA 92093-0424; 
akritsuk@ucsd.edu, 
mnorman@cosmos.ucsd.edu}

\begin{abstract}
We employ a generalization of the She \& L\'ev\^eque model to study velocity 
scaling relations based on our simulations of thermal instability--induced 
turbulence.
Being a by-product of the interstellar phase transition, such multiphase 
turbulence tends to be more intermittent than compressible isothermal
turbulence.
Due to radiative cooling, which promotes nonlinear instabilities in supersonic 
flows, the Hausdorff dimension of the most singular dissipative structures, 
$D$, can be as high as 2.3, while in supersonic isothermal turbulence 
$D$ is limited by shock dissipation to $D\leqslant2$.
We also show that single-phase velocity statistics carry only incomplete
information on the turbulent cascade in a multiphase medium.
We briefly discuss the possible implications of these results on the 
hierarchical structure of molecular clouds and on star formation.
\end{abstract}

\keywords{hydrodynamics --- instabilities --- ISM: structure --- turbulence}

\footnotetext[1]{Also Sobolev Astronomical Institute, 
St. Petersburg State University, Bibliotechnaya Pl. 2, Stary Peterhof, 198504
St. Petersburg, Russia} 

\section{Introduction}
Interstellar turbulence is believed to be neither incompressible
nor isotropic nor homogeneous.
Yet the observed velocity scaling relations in quite a variety of 
environments \citep{larson79,larson81,armstrong..95} are surprisingly 
close to those predicted by Kolmogorov
(1941a, hereafter K41).
At the same time, turbulence in molecular clouds appears to be quite different
in terms of velocity correlations, in some cases, with substantially 
super-Kolmogorov power
indices for both first-order velocity structure functions and velocity
power spectra \citep{brunt.02}.
Interstellar turbulence is one of the key ingredients of modern theories of 
star formation. 
It appears to be the major shaping agent for the mass 
distribution of prestellar dense cores in star forming molecular clouds and, 
ultimately, it could determine the stellar initial mass function 
\citep{padoan.02}.
Therefore, it is important to understand what makes it so similar to
incompressible Kolmogorov turbulence and when, where, and to what extent one
should expect to see significant deviations of observed scaling relations from 
the predictions of K41 theory.

As is known, dissipation controls the scaling properties of turbulence 
through intermittency.
Intermittency corrections to K41 theory for incompressible turbulence, in 
which the most singular dissipative structures are one-dimensional
vortex filaments \citep[hereafter SL94]{she.94}, differ from those for 
compressible supersonic turbulence, in which the structures are 
two-dimensional shock fronts \citep[hereafter B02]{boldyrev02}.
Direct numerical simulations of compressible isothermal turbulence
support this heuristic theory,
demonstrating that the dimension $D$ of the 
dissipative structures depends on the turbulent Mach number and 
changes from $D=1$ for subsonic to $D=2$ for highly supersonic 
flows \citep{padoan03}.

Obviously, the nature of dissipative processes in the interstellar medium (ISM)
is distinct from that in most laboratory environments as well as in 
numerical simulations that assume ideal (magneto)hydrodynamics.
What would the dissipative structures look like in the turbulent ISM
where physical conditions are largely determined by competing cooling and 
heating processes?
The shape of the net cooling function is known to be responsible for thermal 
instability (TI; Field 1965) that affects the ISM 
hydrodynamics in a number of ways, from creating multiple thermal 
phases coexisting at constant pressure \citep{pikel'ner68,field..69} to 
promoting nonlinear instabilities in cold shock-bounded slabs 
\citep{vishniac94,blondin.96}.
How effective are volumetric energy sources in shaping the velocity scalings 
in intermittent interstellar turbulence?
We address this question here by means of three-dimensional numerical 
simulations of decaying hydrodynamic turbulence initiated by rapid thermally 
unstable radiative cooling. 
In this model, turbulence is a by-product of the phase transition in the hot 
gas that leads to the formation of a two-phase medium.
This Letter is the third in our series on turbulence in the multiphase 
ISM \citep{kritsuk.02a,kritsuk.02b}.
We refer the reader to our previous papers for the simulation details 
that are not covered here.

\begin{figure*}
\epsscale{2}
\centerline{\plotone{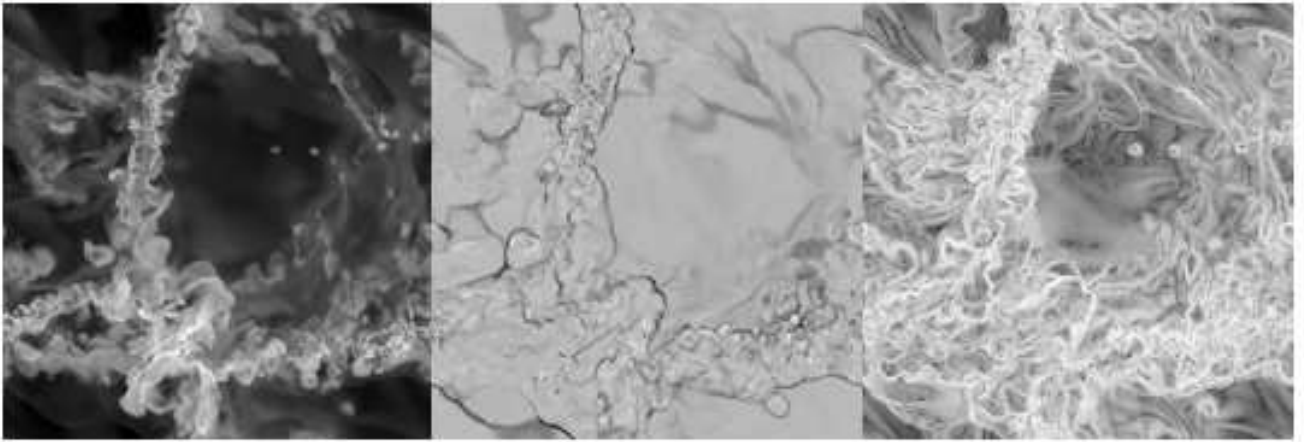}}
\caption{Snapshots of gas density ($\log{\rho}$, {\em left} panel), 
velocity divergence ($\nabla\cdot u$, {\em middle}), and 
vorticity magnitude [$\log(\nabla\times u)^2$, {\em right}], 
for a randomly chosen slice through the computational volume of $256^3$
zones at $t=0.12$~Myr. With a ``standard'' linear gray-scale color map,
dense structures in the left panel are shown as scrambled bright 
filaments, accretion shocks in the middle as sharp dark rims framing the 
cross-section of the density ``walls'', and vortex filaments in the right 
panel 
as delicate freshly cooked Capelli d'Angelo covering the overdense regions 
and extending into underdense voids where the Kolmogorov cascade has not 
quite fully developed yet.}
\vspace{-0.5cm}
\end{figure*}

\section{Model description}
We start the simulation with a periodic box of 5~pc on 
a side filled with hot ($2\times10^6$~K) thermally unstable gas initially 
far from thermal equilibrium (only 2\% of radiative cooling is compensated 
for by a constant uniform volumetric heating source).
Small isobaric random Gaussian density perturbations (power
spectrum index -3) are imposed to initiate TI while initial velocities are
set to zero.
We follow the formation of a turbulent two-phase medium by solving
the equations of ideal gas dynamics (eqs. [6]-[9] in Field 1965), 
assuming zero conductivity, with the piecewise parabolic method
of \citet{colella.84} on a grid of $256^3$ zones. 
The equilibrium cooling function we use describes radiative energy losses
for solar metallicity gas at temperatures $T\in[10,10^8]$~K.

With an initial gas density of 1~cm$^{-3}$, the timescales for cooling and TI
are of the order of $0.1$~Myr. 
The evolution begins with the isobaric linear growth of the density 
perturbations
for $\sim 0.05$~Myr followed by the formation of thermal pancakes --- thin 
two-dimensional cool dense structures emerging on
caustics of the velocity field \citep{sasorov88}.
As the mean gas temperature drops and larger scale modes enter the nonlinear 
regime, a small-scale network of thermal pancakes that formed earlier 
in overdense regions collapses into larger scale accretion shock-bounded 
cold dense structures.
At the end of this {\em rapid cooling} stage, the gas temperature $T\in[30,\; 
3\times10^5]$~K and the flow becomes highly 
supersonic with mass-weighted rms Mach number $\langle M\rangle_{\rho}\gsim15$.
(A simulation initiated with a flat ``white-noise'' spectrum of density 
perturbations returned a substantially lower value of 
$\langle M\rangle_{\rho}\lsim2$, highlighting the importance of large-scale 
modes of TI for generation of supersonic flows.)
Barotropic instabilities, shocks, and nonlinear thin shell instabilities 
\citep{vishniac94} coupled
with TI effectively generate vorticity within the emerging cold dense phase.
As a result, large-scale thermal pancakes \citep{meerson.87} arise turbulent.
Figure~1 shows shock-bounded ``walls'' and underdense ``voids'' forming
during the early {\em turbulent relaxation} period.
The topology of dense structures in Figure~1 is far richer than that 
of two-dimensional shock fronts in supersonic ideal gas turbulence.
This snapshot taken at $t=0.12$~Myr corresponds to the peak of the
volume-weighted rms Mach number, $\langle M\rangle_V\approx3$.
When cooling is present in nonequilibrium supersonic flows like this, 
thin high-density shells get folded by instabilities into sheets of 
finite thickness with $D>2$.

As turbulent relaxation proceeds, gas settles to thermal equilibrium and
thermal phases get separated with about $18$\% of the volume filled 
with the cold phase and $\sim40$\% with the warm phase by $t=0.56$~Myr.
As turbulence decays further, filling factors for both stable phases
slightly grow at the expense of the intermediate thermally unstable regime.
Since this two-phase medium is quasi-isobaric 
($\Delta p/p\approx 6\ll\Delta\rho/\rho\approx1000$) 
and since velocities inherited from the initial ``violent relaxation''
do not correlate with density, the Mach number depends 
on the gas density roughly as $M\propto\rho^{1/2}$ (see Fig.~2).
This means that there are two different regimes of turbulence coexisting 
in such a two-phase medium: a subsonic regime with $M\sim0.4$ 
in the space-filling warm phase and a supersonic one $M\sim1.5$ in the cold 
phase.
The Mach number--density relation is a distinct feature of turbulence in 
a two-phase medium.
It is quite natural that velocity scaling relations in this case will not 
be identical to those for isothermal compressible turbulence.
It is also clear that reconstruction of these relations based on observations
of any single phase in a multiphase medium would not be an easy task.
We will further illustrate these ideas in the following section by formally 
applying statistical analysis methods developed for turbulence research to our
simulation data.

\section{Velocity statistics}
In order to follow the time evolution of the turbulent velocity field
across different scales, we computed 
three-dimensional velocity power spectra for four snapshots taken at
$t=0.12$, 0.56, 1.5, and 2.5~Myr (Fig.~2).
We also decomposed the velocity field into a divergence-free {\em solenoidal}
component $u_s$ ($\nabla\cdot u_s=0$, dotted lines) and a 
curl-free {\em dilatational} component $u_d$ ($\nabla\times u_d=0$, 
dashed lines), such that $u=u_s+u_d$.
As turbulence develops and decays, the fractions of power contained in the 
solenoidal and dilatational components vary as a function of time and scale.
At $t=0.12$~Myr, the power spectrum is dominated by dilatational 
modes on large scales.
A ``wave'' of solenoidal modes generated by nonlinear instabilities 
propagates from small to large scales and overlaps with dilatational modes 
as turbulence develops.
Earlier, at $t=0.07$~Myr, the dilatational spectrum follows a nearly perfect
$k^{-2}$ law for the whole range of wavenumbers, characteristic of 
Burgers turbulence accompanying the formation of a multiscale network of 
thermal pancakes.
At that moment, the contribution of solenoidal modes is negligibly small.
\begin{figure*}
\vspace{-0.2cm}
\centerline{\plottwo{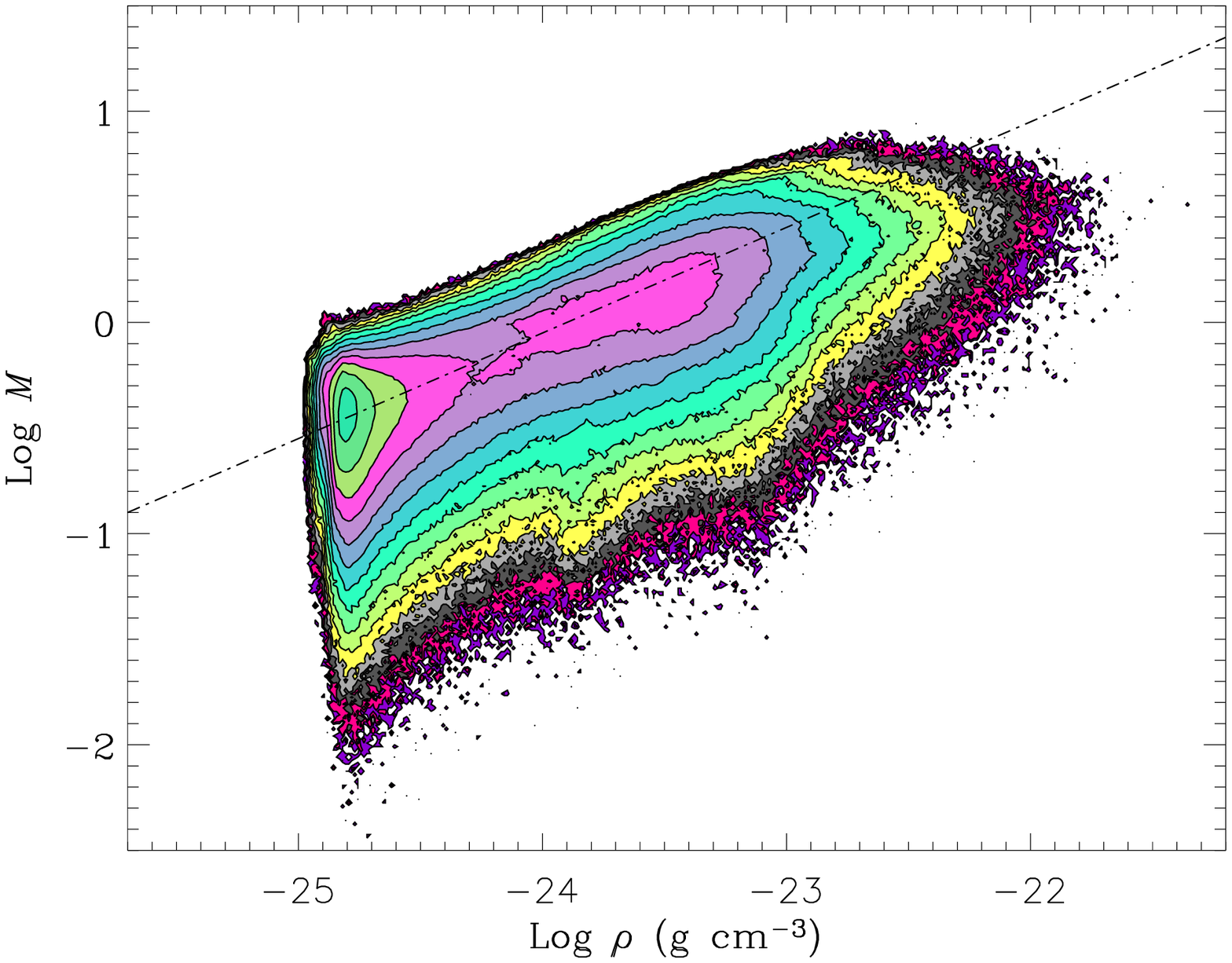}{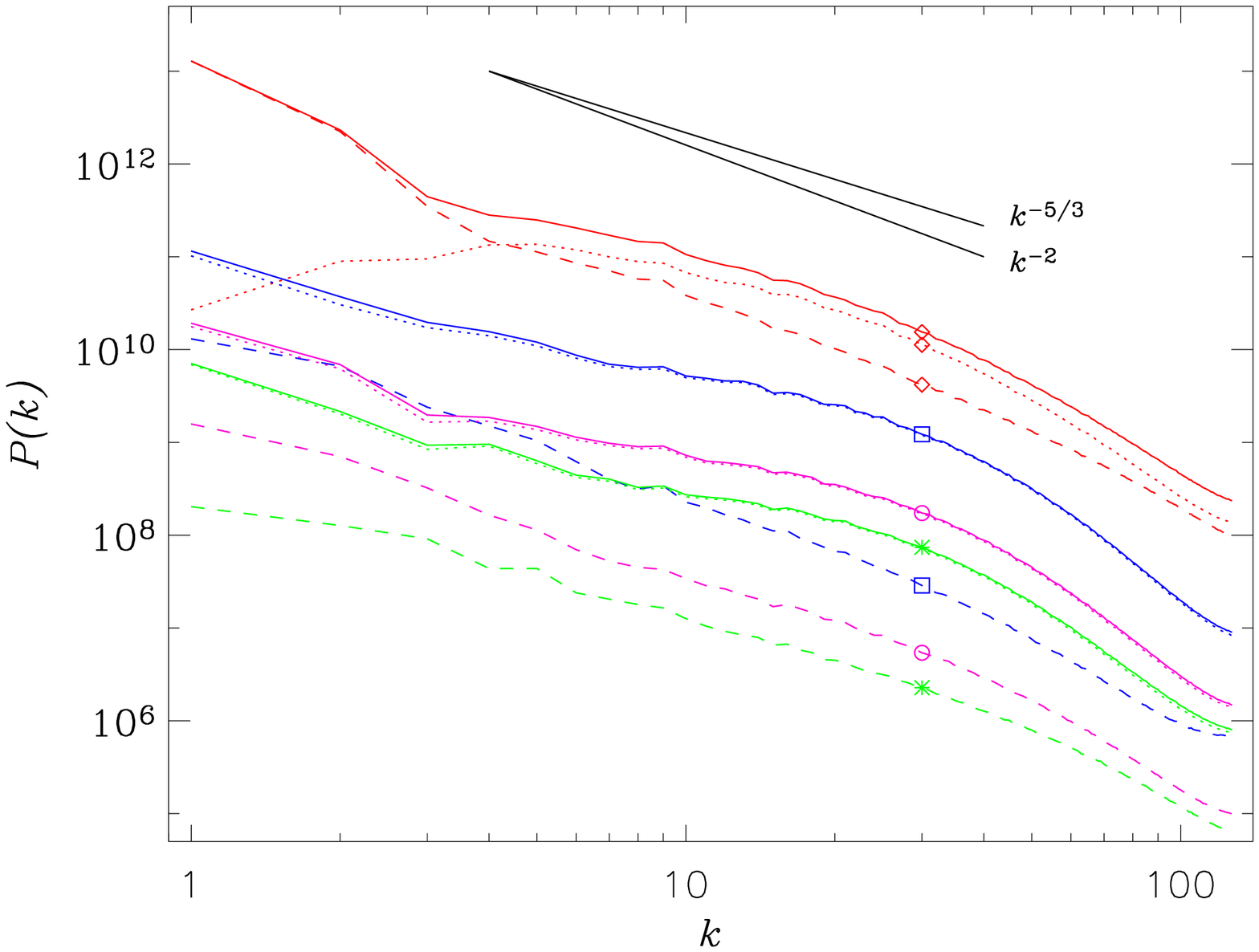}}
\vspace{-0.1cm}
\caption{{\em Left panel}: Scatter plot of Mach number vs. gas density
at $t=0.56$~Myr; contour levels are consecutive powers of 2, and the 
dash-dotted line follows the relation $M\propto \rho^{1/2}$
for isobaric gas with uncorrelated density and velocity.
{\em Right panel}: Three-dimensional velocity power spectra 
({\em solid lines}) 
for $t=0.12$ ({\em diamond}), 0.56 ({\em square}), 1.5 ({\em circle}), and 
2.5~Myr ({\em asterisk}) decomposed into solenoidal 
({\em dotted lines}) and dilatational 
({\em dashed lines}) components. Also shown are the 
velocity power spectra for Kolmogorov (-5/3) and
Burgers (-2) turbulence.}
\vspace{-0.5cm}
\end{figure*}
By $t=0.56$~Myr, solenoidal modes dominate on virtually all scales since
dilatational modes decay much faster than solenoidal in an unforced 
regime that follows the initial phase transition.
At $t=1.5$~Myr, the latter contribute only up to a few percent, making the 
dotted and solid lines in Figure~2 nearly indistinguishable.
The dependence of power in solenoidal versus dilatational modes on the rms 
Mach number of the flow in our simulations is broadly consistent with the
results of \citet{boldyrev..02b} for driven isothermal MHD turbulence.

To study scaling properties of the decaying turbulent cascade, we compute 
longitudinal and transverse velocity structure functions (SFs) 
\begin{equation}
S_p(l)=\langle|u(x)-u(x+l)|^p\rangle\propto l^{\zeta_p}
\label{SF_def}
\end{equation}
\citep{monin.71}
and exploit extended self-similarity \citep{benzi.....93,benzi....96} 
to obtain estimates for scaling exponents $\zeta_p$ relative to the third-order
exponent: $S_p\propto S_3^{\zeta_p/\zeta_3}$.
These {\em relative} scaling exponents may be more universal than $\zeta_p$ 
themselves since it is natural to use a generalized scale $\ell(l,\eta)$ 
instead of the resolution scale $l$ for the investigation
of the scaling properties of turbulence \citep{dubrulle94}.
A value of $\zeta_3$ is still required to recover the absolute values of the 
exponents $\zeta_p$, e.g., to get an estimate for the velocity power spectrum 
scaling in the inertial range $P(k)\propto k^{-1-\zeta_2}$.
A rigorous result $\zeta_3=1$ holds for incompressible Navier-Stokes 
turbulence (\citet{kolmogorov41b} ``$4/5$'' law) and for incompressible MHD 
\citep{politano.98}.\footnote{Strictly speaking, $\zeta_3=1$
is proved only for the longitudinal structure function, $S_3$, 
\citep{kolmogorov41b} and for certain mixed structure functions 
\citep{politano.98}; in both cases the absolute value of the velocity 
difference is {\em not} taken. However, it is believed that $\zeta_3=1$ 
holds in eq. (\ref{SF_def}) as well.}
Numerical simulations support the same result for supersonic compressible 
hydrodynamics \citep[hereafter PPW02]{porter..02}, but it is unclear whether 
this remains true in the presence of a generic volumetric energy source.

A three-parameter model developed by \citet{she.94} and generalized by
\citet{dubrulle94} represents the turbulent energy cascade as an infinitely 
divisible log-Poisson process and relates the dimensionality of the most 
dissipative structures $D$ to the relative scaling exponents:
\begin{equation}
\frac{\zeta_p}{\zeta_3}=(1-\Delta)\Theta p +
\frac{\Delta}{1-\beta}(1-\beta^{\Theta p}),\label{d94}
\end{equation}
where $\beta\equiv1-\Delta/(3-D)$ measures the ``degree of nonintermittency'' 
of energy dissipation, $\beta\in(0,1)$.
The other two parameters represent nonintermittent
scalings for the velocity difference, $u_l\propto l^\Theta$, and
for the eddy turnover time scale, $t_l\propto l^{\Delta}$.
For the Kolmogorov cascade, $\Theta=\frac{1}{3}$ and $\Delta=\frac{2}{3}$.
In the limit $\beta\to1$, the system is nonintermittent, and the K41 relation 
can readily be recovered.
Since $\beta\to0$ as $D\to2\frac{1}{3}$, the model is only consistent with 
$D<2\frac{1}{3}$.
Vortex filaments in incompressible turbulence are characterized by $D=1$
with which eq. (\ref{d94}) reduces to SL94 formula.
In a series of numerical experiments on isothermal compressible turbulence
with Mach numbers $0.62\leqslant M\leqslant10$, \citet{padoan03} 
were able to trace a smooth transition from scalings that can be
approximated by equation (\ref{d94}) with $D=1$ to those with $D=2$. 
Their result is consistent with the suggestion that velocity SF scaling 
for compressible turbulence can be described by the She-L\'ev\^eque 
formalism for any value of the Mach number of the flow, with only one 
parameter, $D$, varying as a function of $M$.

What kind of scaling can we expect for our model with a ``realistic'' 
nonisothermal scale-dependent effective equation of state determined 
by cooling and heating? 
How will it evolve as turbulence decays and the two phases settle into
thermal and pressure equilibrium?

We computed velocity SFs up to the sixth order using the so-called 
XYZ method (see, e.g., PPW02).
Relative scaling exponents for the same four snapshots that we used for 
power spectra are shown in Figure 3. 
We consider the difference between exponents for longitudinal and transverse 
SFs as a good quality indicator for the derived scalings (larger 
symbols show larger statistical uncertainties).
We attribute this difference
to poor statistics and to small deviations from full isotropy in
our numerical model due to an insufficiently large value of the Reynolds 
number and the
slightly anisotropic initial forcing on scales approaching the box size.

\begin{figure*}
\vspace{-0.2cm}
\centerline{\plottwo{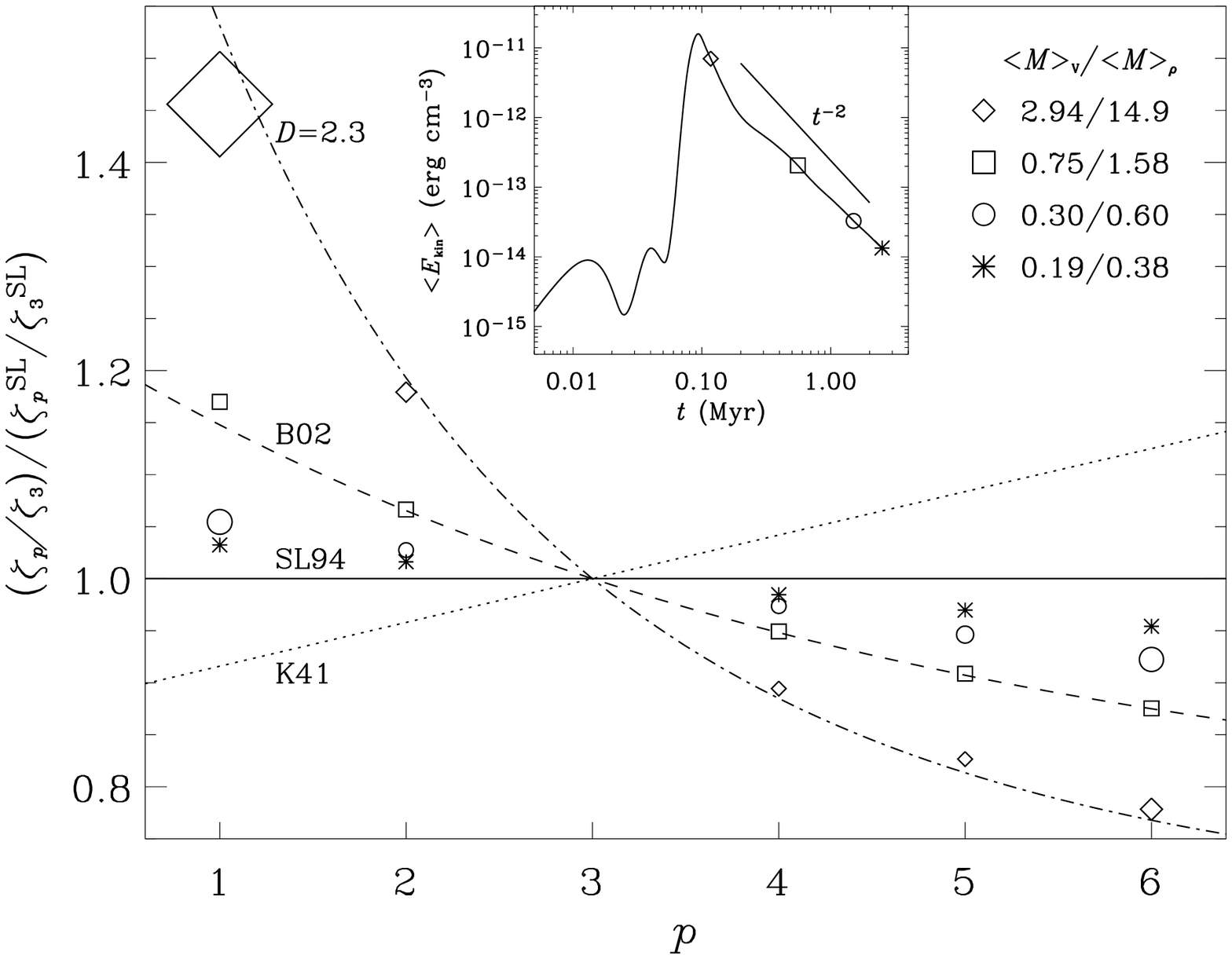}{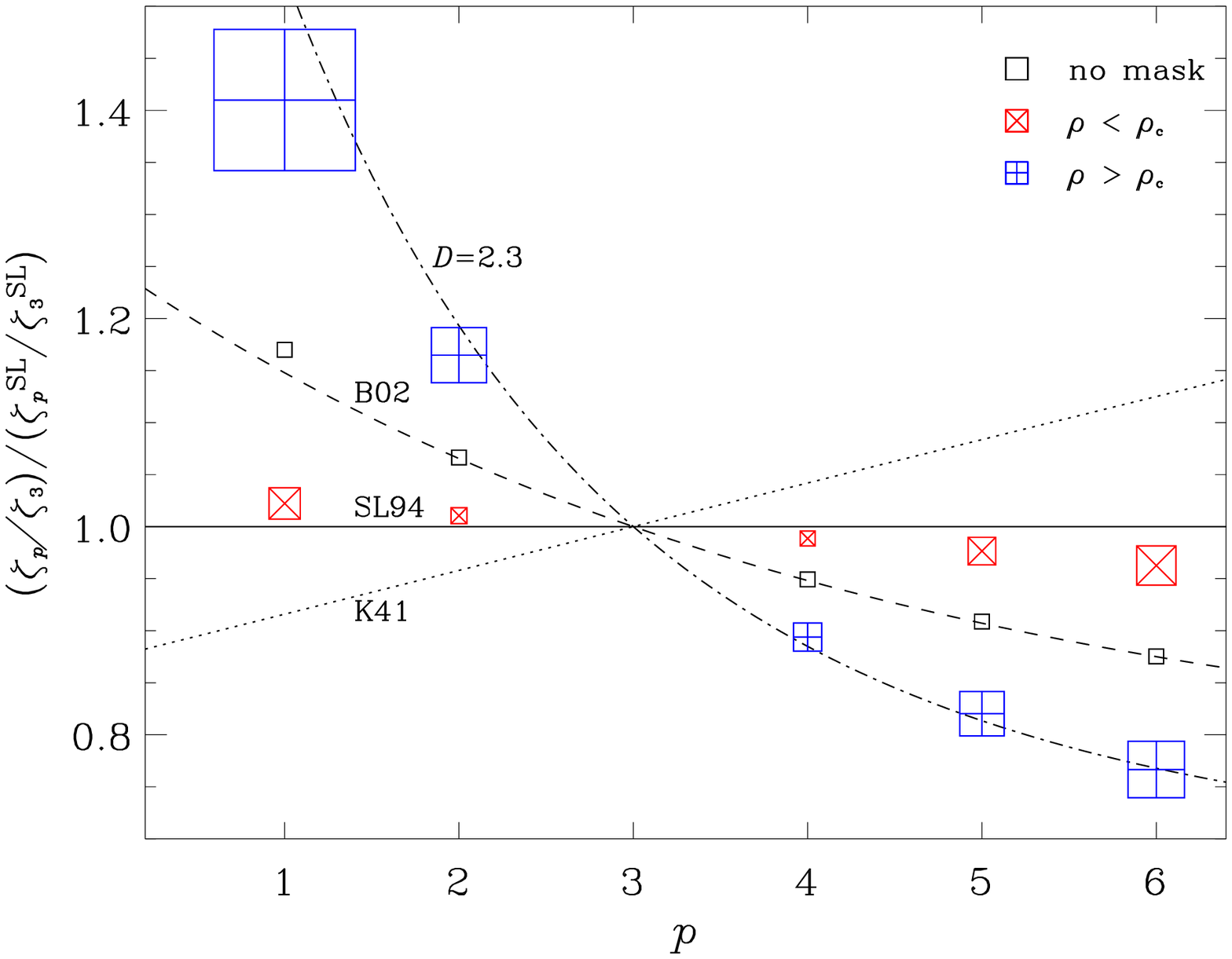}}
\vspace{-0.1cm}
\caption{Scaling exponents for velocity structure functions of the orders
 1--6 normalized by the SL94 model ({\em horizontal solid line}); 
other models include K41 ({\em dotted line}), B02 ({\em dashed line}), 
and eq. (\ref{d94}) with a fractal dimension $D=2.3$ 
({\em dash-dotted line}). 
Symbol sizes show the difference between the longitudinal and
transverse structure functions with a floor value.
{\em Left panel}: Four snapshots illustrating the evolution of the scaling.
The insert shows the kinetic energy as a function of time and timing for 
the snapshots:
0.12 ({\em diamonds}),
0.56 ({\em squares}),
1.5 ({\em circles}), and
2.5~Myr ({\em asterisks}).
The legend in the upper right-hand corner gives volume- and 
mass-weighted rms Mach
numbers for the snapshots.
{\em Right panel}: Scaling exponents for a snapshot at $t=0.56$~Myr.
The open squares are the same as in the left panel; 
the crossed squares show the exponents obtained with density masks 
$\rho\gtrless\rho_c=10^{-24.3}$~g~cm$^{-3}$ and probe velocity
correlations in the cold and warm phases separately.}
\vspace{-0.5cm}
\end{figure*}
 
Our results generally agree with those of \citet{padoan03} in the
sense that the exponents in Figure 3 can be well described by equation 
(\ref{d94})
and, as the turbulence decays and becomes less supersonic, $D\to1$.
However, for a given turbulent Mach number, the value of $D$ appears to be 
larger in our multiphase case than in the isothermal turbulence.
This could potentially be caused by the possibility that $\zeta_3<1$ in 
our case; higher resolution simulations would help to test this option.
It is tempting to note that our first snapshot (diamonds), which corresponds
to the most supersonic regime with no clear distinction between the thermal 
phases, demonstrated the maximal value of $D$ for this run which is
approaching the upper bound of $2\frac{1}{3}$ from below.
Curiously enough, this value coincides with the observationally determined
fractal dimension of molecular clouds $D=2.3\pm0.3$ \citep{elmegreen.96}.
We refer to Boldyrev et al. (2002b) for a relation between density and 
velocity correlators.
It seems, however, that it is not the limited numerical resolution,
but rather the lack of cooling-related instabilities that prevents 
dense structures from folding into a fractal distribution with dimensionality 
$D\gsim2$ in simulations of isothermal turbulence (cf. Boldyrev, Nordlund, \& 
Padoan 2002a).
We will address the effects of magnetic fields and gas self-gravity 
on the dimension of the dissipative structures in a subsequent paper.

Observational diagnostics of turbulent velocity fluctuations in the ISM
are usually based on phase-specific spectral line transitions.
Since we model the formation of a turbulent two-phase medium, it is natural 
to compute SFs by measuring the velocity differences within a single phase 
and then to compare the results for both phases.
We introduced two masks based on a gas density value of 
$\rho_c=10^{-24.3}$~g~cm$^{-3}$ and obtained velocity SFs separately for zones
where the gas density is higher or lower than the threshold $\rho_c$, which is
close to the lower bound of a thermally unstable regime (for gas in thermal 
equilibrium) and, thus, separates the warm stable phase 
from cold ``clouds'' and unstable gas.
At $t=0.56$~Myr, the filling factors for both density regimes are about 50\%.

The scaling exponents for velocity SFs of low- and high-density gas are shown
in Figure 3.
The statistics for both density regimes are relatively poor 
as is indicated by the size of the crossed symbols.
They are good enough, however, to conclude that the cold-phase exponents 
show significantly larger departures from K41 scalings than do the complete 
statistics unaffected by masking.
The crossed symbols representing the warm phase instead demonstrate more 
K41-like scalings.
Both transverse and longitudinal SFs follow the trend outlined above, but
the exponents for transverse SFs show consistently smaller departures from
the unbiased exponents.
We conclude that the velocity power spectrum index $n_c$
based on observations of the cold molecular phase would instead be an upper
estimate for the actual index $n=1+\zeta_2$, while $n_w$ determined from 
the warm gas velocities would only give a lower bound for $n$.

\section{Conclusions}

We considered a simple model for TI-induced decaying hydrodynamic turbulence 
in a radiatively cooling medium with a low constant heating rate.
A formal application of the \citet{she.94} model to the simulated velocity 
fields consistently showed that cooling-related instabilities make turbulence 
in multiphase gas more intermittent than conventional compressible supersonic 
turbulence.
This result could have interesting implications for the recently developed
turbulent fragmentation branch of star formation theory.
Finally, we demonstrated that phase-specific observational velocity statistics 
for multiphase media can be affected by the correlation of a turbulent Mach 
number with gas density.

\acknowledgments

We acknowledge useful discussions with Paolo Padoan.
This work was partially supported by NRAC allocation MCA098020
and utilized computing resources provided by the San Diego Supercomputer 
Center.

\end{document}